\documentclass[ 
reprint,
 amsmath,amssymb,
 aps,showpacs,superscriptaddress
]{revtex4-1}

\usepackage{graphicx}
\usepackage{dcolumn}
\usepackage{bm}
\usepackage{subfig}
\usepackage{cancel}
\usepackage{mathtools} 
\usepackage{upgreek}

\usepackage{color}
\usepackage{enumitem}
\newcommand{\beqar}{\begin{eqnarray}}
\newcommand{\eeqar}{\end{eqnarray}}

\newcommand{\bcen}{\begin{center}}
\newcommand{\ecen}{\end{center}}

\newcommand{\eps}{\varepsilon}
\newcommand{\lam}{\lambda}

\newcommand{\ket}[1]{\left| #1 \right>}

\newcommand{\f}[2]{\frac{#1}{#2}}
\renewcommand{\b}[1]{\left({#1}\right)}
\renewcommand{\v}[1]{\vec{#1}}
\newcommand{\pd}[2]{\frac {\partial #1}{\partial #2}}

\renewcommand{\sb}[1]{\left[{#1}\right]}
\newcommand{\mean}[1]{\langle {#1} \rangle}
\newcommand{\infint}{\int_{-\infty}^{\infty}}
\newcommand{\ra}{\rightarrow}

\begin{document}

\preprint{APS/123-QED}
\title{Fast Route to Thermalization}

\affiliation{The Institute of Chemistry, The Hebrew University of Jerusalem, Jerusalem 9190401, Israel}%
\affiliation{Department of Physical Chemistry, University of the Basque Country UPV/EHU, Apdo 644, Bilbao, Spain}
\affiliation{Kavli Institute for Theoretical Physics, University of California, Santa Barbara, CA 93106, USA}

\author{Roie Dann}
\email{roie.dann@mail.huji.ac.il}
\affiliation{The Institute of Chemistry, The Hebrew University of Jerusalem, Jerusalem 9190401, Israel}%
\affiliation{Kavli Institute for Theoretical Physics, University of California, Santa Barbara, CA 93106, USA}
\author{Ander Tobalina}
\email{ander.tobalina@ehu.eus}
\affiliation{Department of Physical Chemistry, University of the Basque Country UPV/EHU, Apdo 644, Bilbao, Spain}
\affiliation{Kavli Institute for Theoretical Physics, University of California, Santa Barbara, CA 93106, USA}
\author{Ronnie Kosloff}%
\email{kosloff1948@gmail.com}
\affiliation{The Institute of Chemistry, The Hebrew University of Jerusalem, Jerusalem 9190401, Israel}%
\affiliation{Kavli Institute for Theoretical Physics, University of California, Santa Barbara, CA 93106, USA}

\date{\today}
\begin{abstract}
    We present a control scheme for quantum systems coupled to a thermal bath. We demonstrate state-to-state control between two Gibbs states. This scheme can be used to accelerate thermalization and cool the open system. Starting from a microscopic description, we derive the reduced system dynamics, leading to a non-adiabatic master equation. The equation contains non-trivial effects due to the non-adiabatic driving and bath interaction. These special features enable controlling the open system and accelerating the entropy changes. For a two-level system model, we obtain a general solution and introduce a reverse-engineering scheme for control. The control problem is analyzed in the context of the theory of quantum control and the accompanying thermodynamic cost.
\end{abstract}

                    
\maketitle

\section{Introduction}
\label{sec:intro}

Thermalization, the dynamical approach of a system towards  thermal equilibrium,
is of the utmost importance in contemporary physics. Typically, thermalization occurs once a system interacts with a large thermal reservoir of a defined temperature.
In the quantum world, this dynamical process is embedded in
the theory of open quantum systems and is consistent with the laws of thermodynamics \cite{alicki2018introduction}.

In present-day quantum physics, thermalization is commonly employed to initialize a system state 
before further manipulations are preformed. For example, the  initialization step in circuit quantum computing is a thermalization step
and adiabatic quantum computing relies completely on thermalization to carry out a computation protocol \cite{Brooks:2013aa,Campbell:2017aa,Albash:2017ab}. In these examples,  experimenters utilize the system-bath interaction to reduce the system temperature, suppressing the effective number of quantum states.
Time-domain pump-probe spectroscopy relies on thermalization to close the loop of repeated experiments \cite{mukamel1995principles}. 
Moreover,  thermalization is  a vital step
in quantum heat engines, and its timescale dictates the power output of the device \cite{alicki1979quantum,geva1992quantum,kosloff2017quantum,feldmann2003quantum,von2019spin,pekola2019supremacy}.

Typically, thermalization is treated as a passive process.  An interaction between a non-equilibrium system and the environment is introduced, and the thermal temperature of the environment and the system-bath coupling
dictate the typical relaxation timescale. Moreover, after the initial thermalization step, the experimenter usually invests a considerable effort in isolating the system
from environmental effects to  attenuate the relaxation rate.  In contrast to these approaches, the present study aims to harvest the environmental interactions to actively control the system and accelerate the thermalization.

Quantum control relies on the interference of pathways to achieve the control objective.
Such interference requires a non-stationary superposition of states, i.e., quantum coherence.
Commonly, coupling to a thermal bath causes decoherence, which tends to suppress the ability to control.
Conversely, in the presented scheme, this interplay between coherence generation by the controller and decoherence by
the bath will be shown to enable control.

We addressed this issue in a short letter \cite{dann2019shortcut}: {\em Shortcut to Equilibrium }(STE), where we described a 
procedure inducing swift thermalization by a rapid change in the system Hamiltonian.
The process was demonstrated for a parametric harmonic oscillator coupled to a bath, and later combined with unitary strokes to study quantum analogues of the Carnot cycle at finite time \cite{dann2019carnot}. 
In the current paper, we present a comprehensive account of the fast thermalization process.
We emphasize the relation between quantum control and quantum thermodynamics, employing the thermalization of a qubit as our primary example. This allows for a detailed thermodynamics account of the related cost of rapid thermalization. We further extend the control scheme to include an initial non-equilibrium Gibbs state and a protocol that cools the system below the temperature of the bath.  In addition, we elaborate on the connection to quantum coherent control.

The paper is organized as follows: we set the control framework and objective in Sec. \ref{sec:control objective}. Sec. \ref{sec:driven dynamics} presents a first principle analysis of the open system dynamics. We begin by solving for the propagator of the isolated system utilizing the inertial theorem, then include a weak interaction between the system and the bath, and present a  brief construction of the Non-Adiabatic Master Equation (NAME). Following the dynamical construction, Sec. \ref{sec:control} analyzes the task of rapid thermalization within the framework of Quantum Control theory. Section \ref{sec:STE for TLS} constructs the dynamical description of a two-level system coupled to a thermal bath, demonstrating the general derivation. Next, we solve the two-level system dynamics and discuss its properties. Following the solution, we present a reverse-engineering control scheme to obtain a STE protocol that induces rapid thermalization. Generalizations of the scheme are considered, including an initial non-equilibrium Gibbs state and a target state which is colder than the bath. Section \ref{sec:Results and Discussion} is devoted to results and discussion. We conclude in Sec. \ref{sec:Conclusions} with a summary, and discuss relevant implications, and future prospects.

\section{Control objective and framework}
\label{sec:control objective}
We aim to find a rapid thermalization protocol of a subsystem $S$ coupled to a bath. This can be generalized to state-to-state control from an initial Gibbs state $\rho_S^i=\frac{1}{Z_{S,i}}\exp\b{-\beta_i \hat H_S\b{t_i}}$ with Hamiltonian $\hat{H}_S\b{t_i}=\hat{H}_S^i$ and inverse temperature $\beta_i$, toward the target state $\hat \rho_S^f=\frac{1}{Z_{S,f}}\exp(-\beta_f \hat H_S\b{t_f})$ with a final Hamiltonian $\hat{H}_S\b{t_f}=\hat{H}_S^f$ and inverse temperature $\beta_f$.
This objective falls into the category of state-to-state control of open quantum systems \cite{koch2016controlling}.

The control framework lies in a composite Hilbert space partitioned into a system and a bath. The combined system's evolution is governed by the total Hamiltonian 
\begin{equation}
    \hat{H}\b t =\hat{H}_S \b t +\hat{H}_B+\hat{H}_I~~,
    \label{eq:Ham_tot}
\end{equation}
where $\hat{H}_S\b t$ is the time-dependent system Hamiltonian which includes the external driving. $\hat{H}_B$ is the bath Hamiltonian and $\hat{H}_I$ is the system-bath interaction term. To proceed, we make three basic assumptions:
\begin{enumerate}[label=(\alph*)]
    \item  Only the system Hamiltonian is controllable, via external driving;
    \item The bath is sufficiently large to maintain a thermal state at all times;
    \item  The system bath coupling is weak (and given a priori).
\end{enumerate}

A solution for the control problem consists of two major steps: first, obtaining accurate reduced dynamical equations of motion for the system variables; second, finding a framework where the control problem can be solved parametrically, based on the equations of motion.

\section{Driven dynamics of the open system}
\label{sec:driven dynamics}
To achieve control, we require a dynamical equation of motion for a non-adiabatically driven system coupled to a bath. We employ a derivation based on first principles which complies with thermodynamic principles. 
The reduced dynamics is then formulated by a completely positive trace-preserving map, generated the by Gorini-Kossakowski-Lindblad-Sudarshan (GKLS) master equation \cite{gorini1976completely,lindblad1976generators}:
\begin{equation}
 \frac{d}{dt}{{\hat\rho_S\b t} }=   -\frac{i}{\hbar}[{\hat{H}_S\b t },{ \hat{\rho}_S\b t}] +{\mathcal{L}_D\b t}{\hat{\rho}_S\b t}~~,
  \label{eq:lindblad}
\end{equation}
where $\cal{L}_D$ is the dissipative part.
The main issue lies in the fact that the dissipative part in the equations of motion depends on the system Hamiltonian and is therefore influenced by the driving. Neglecting this contribution leads to inconsistencies with thermodynamics \cite{alicki2018introduction}. 

\subsection{Inertial solution of the free dynamics}
\label{subsec:Inertial_theorem_solution}
An accurate description of the open system dynamics requires first, an explicit solution of the free dynamics ${\cal U}_S\b t$. Obtaining such a solution is difficult when the free Hamiltonian does not commute at different times $\sb{\hat{H}_S\b t,\hat{H}_S\b{t'}}\neq  0$. In this case,
there is no common basis that diagonalizes ${\cal U}_S\b t$. A formal solution is given by a time-ordering procedure,
\begin{equation}
    {\cal U}_S\b t = \overleftarrow{\cal T} \exp\b{ -i \int_0^t [\hat H_S\b{t'}, \bullet ] dt' }~~,
    \label{eq:torder}
\end{equation}
where $\overleftarrow{\cal{T}}$ is the chronological time-ordering operator. 
However, this solution contains an infinite sum, and is therefore impractical for our purposes.  
An explicit solution requires bypassing the time-ordering obstacle. 
This is achieved by utilizing the inertial theorem and solution. The inertial solution is a consequence of the theorem; it approximates the system dynamics under the condition of slow acceleration of the drive \cite{dann2018inertial}.

The proof of this property is established in the framework of the Liouville space representation and assumes a system with a closed Lie algebra.
 Liouville space is a Hilbert space of system operators, embedded with an inner product $\b{\hat{A},\hat{B}}\equiv\text{tr}\b{\hat{A}^\dagger\hat{B}}$ \cite{mukamel1995principles,petrosky1997liouville,am2015three}. In this space, the quantum system is represented in terms of a finite operator basis. These operators form a vector $\v v $ in Liouville space. The operator dynamics are calculated using the Heisenberg equation, and for a closed operator algebra, the dynamical equations are cast in a matrix vector form
\begin{equation}
\f{d\v v \b t}{dt} =-i{\cal{M}}\b t\v v \b t~~.    
\label{eq:dynamics}
\end{equation}
 Here, ${\cal{M}}\b t$ is the dynamical generator in Liouville space. For a varying Hamiltonian the generator has an explicit time-dependence.

The inertial theorem relies on a priori decomposition of ${\cal{M}}\b t$ into a time-dependent scalar function $\Omega\b t$ and a constant matrix $\cal{B}\b{\v{\chi}}$, dependent on constants $\{ \chi_i \}$, written in short notation as $\v{\chi}=\{\chi_1,\chi_2,...,\chi_m \}^T$:
\begin{equation}
{\cal{M}}\b t=\Omega\b t {\cal{B}\b{\v \chi}}~~.    
\label{eq:inertial_decompasition}
\end{equation}
Such a decomposition is achieved by expressing the Liouville dynamics in terms of a suitable time-dependent operators basis $\v v\b t=\{\hat{v}_1\b t,...,\hat{v}_N\b t  \}^T$ and control protocol $\hat{H}_S\b t$.

Combining Eqs. \eqref{eq:dynamics} and \eqref{eq:inertial_decompasition} and diagonalizing ${\cal{B}}$, we find the eigenvectors $\{\v{F}_k \}$ of ${\cal{U}}_S\b t = \exp \b{-i{\cal{B}}\int_0^t \Omega\b{t'}dt'}$. These Liouville vectors define the eigenoperators, $\{\hat{F}_k\}$, of the free propagator, according to the mapping $\hat{F}_k =\sum_{i=1}^N f_i \hat{v}_i$, where $f_i$ and $\hat{v}_i$ are the elements of $\v F_k$ and $\v{v}$. The set of eigenoperators forms a basis in Liouville space and satisfies an eigenvalue-type equation
\begin{equation}
\hat{F}_k \b{\v{\chi},t} =\hat{U}_S^\dagger\b{t}\hat{F}_k \b{\v \chi,0}\hat{U}_S\b{t} = e^{-i\lambda_k \theta \b t} \hat{F}_k\b{\v \chi,0} ~~,
\label{eq:eigenvalue eq}
\end{equation}
where $\hat{U}_S\b t$ is the free propagator in the Schr\"odinger representation, and $\lambda_k$ is the $k$'th eigenvalue of ${\cal{B}}$ and $\theta \b t =\int_0^t \Omega \b{t'}dt'$.
When decomposition \eqref{eq:inertial_decompasition} is found, for a slowly varying $\{ \chi_i\b t \}$ ($d\chi_i/d\theta\ll 1$), the evolution of the eigenoperators is approximated by the inertial solution
\begin{equation}
\hat{F}_k \b t\equiv\hat{F}_k \b{\v{\chi}\b t,t}=e^{-i\int_{\theta\b 0}^{\theta\b t} {\lambda_k\b{\theta'}d\theta'}}e^{i\phi_k} \hat{F}_k \b{ \v{\chi}\b t,0}
~~. 
\label{eq:gen_inertial_theorem1}
\end{equation}
Here, $\lambda_k\b{\theta\b t}$ are the instantaneous eigenvalues and $\phi_k$ is a geometrical phase \cite{dann2018inertial}. Since $\{\hat{F}_k\}$ form a complete operator basis, under the condition of a slow varying $\{ \chi_i\b t \}$, Eq. \eqref{eq:gen_inertial_theorem1} fully determines the system dynamics. Finally, we introduce the instantaneous diagonalization matrix ${\cal{V}}\b{\v \chi\b t}$ of ${\cal{B}}\b{\v{\chi}\b t}$ to obtain the inertial dynamics of the operators of $\v v \b t$
\begin{multline}
    \hat{v}_i\b t=\sum _{i,k}{\cal{V}}^{-1}_{ik}\hat{F}_k\b t\\=
    \sum_{k}{\cal{V}}_{ik}\b{\v \chi \b t}e^{-i\int_{\theta\b 0}^{\theta \b t} {\lambda_k\b{\theta'}d\theta'}}
    \hat{F}_k\b{\chi\b t,0}\\=
    \sum_{k,j}{\cal{V}}_{ik}\b{\v \chi \b t}e^{-i\int_{\theta\b 0}^{\theta \b t} {\lambda_k\b{\theta'}d\theta'}}
    {\cal{V}}^{-1}_{kj}\b{\v \chi\b t} \hat{v}_j\b{0}~~.
    \label{eq:gen_inertial_theorem}
\end{multline}

\subsection{Construction of the NAME}
\label{subsec:Open_system_dynamics}
 We are now prepared to include the environmental influence and describe the dynamics of  a non-adiabatically driven open quantum system. In this regime, when the typical driving time-scale is comparable to the system Bohr frequencies $\Delta E/\hbar$, the external driving dresses the interaction with the bath. As a result, for fast driving,  the adiabatic master equation \cite{albash2012quantum,sarandy2005adiabatic,yamaguchi2017markovian} is inadequate \cite{dann2018time}. 
 
 To consistently describe the evolution of a non-adiabatically driven open quantum system, we construct the equation of motion utilizing a first principle derivation.
 In the spirit of the Davis construction \cite{davies1974markovian}, we begin with a complete description of the composite system, Eq. \eqref{eq:Ham_tot} and assume weak-system-bath coupling.  The complete dynamics is governed by the Liouville equation, which, within the Born-Markov approximation \cite{breuer2002theory}, leads to the Markovian Quantum Master Equation 
 \begin{equation}
     \f{d}{dt}\tilde{\rho}_S\b t=-\f{1}{\hbar^2}\int_{0}^{\infty} ds\, \text{tr}_B\sb{\tilde{H}_I\b t,\sb{\tilde{H}_I\b{t-s},\tilde{\rho}_S\b t\otimes\hat{\rho}_B}}~~.
     \label{eq:QMME}
 \end{equation}
Here, $\tilde{\rho}_S\b t={\cal{U}}_S\b t\hat{\rho}_S\b t$ is the reduced density matrix in the interaction representation, and  a similar notation is applied for any system operator.

The interaction Hamiltonian can always be written as a sum of separable terms $\hat{H}_I=\sum_n \hat{S}_n\otimes \hat{B}_n$, where $\hat{S}_n$ and $\hat{B}_n$ are system and bath operators.
Expressing the interaction term 
 in terms of the operators $\hat{v}_i$, $\hat{S}_n=\sum_{i} s_{ni}\hat{v}_i\b 0$ and assuming inertial driving, the inertial solution, Eq. \eqref{eq:gen_inertial_theorem}, leads to 
 \begin{multline}
\tilde{H}_I\b t = \sum_{n,k} c_{nk}\b{\v \chi\b t} e^{-i\int_{\theta \b 0}^{\theta\b t} \lam_k\b{\theta'}d\theta'}\\
\times\hat{F}_k\b{\v{\chi}\b t,0}\otimes \tilde{B}_n\b t\\
 = \sum_{n,k} \xi_{nk}\b{\v \chi\b t}e^{-i\Lambda_{nk}\b t}\hat{F}_k\b{\v \chi\b t,0}\otimes \tilde{B}_n\b t~~,
 \label{eq:H_int_gen}
 \end{multline}
with $\Lambda_{n,j}\b t \equiv \int_{\theta\b 0}^{\theta\b t}{\lam_k\b{\theta'}d\theta'}+\eta_{nk}\b t$, where $\eta_{nk}$ and $\xi_{nk}$ are the phase and positive amplitude of $c_{nk}\b{\v \chi\b t}=\sum_i s_{ni}{\cal{V}}^{-1}_{ik}\b{\v \chi\b t}=\xi_{nk}\b te^{-i\eta_{nk}\b t}$, respectively.
Note, that the Schr\"odinger and interaction pictures coincide when the operators are implicitly independent of time, $\tilde{F}_k\b{\v{\chi}\b t,0}=\hat{F}_k\b{ \v\chi\b t,0}$.

 Overall, the interaction term in the interaction representation is given as an expansion of the instantaneous eigenoperators $\hat{F}_k\b{ \v\chi\b t,0}$, depending weakly on time trough $\v \chi \b t $. Now, we substitute Eq. \eqref{eq:H_int_gen} into Eq. \eqref{eq:QMME} and 
expand $\Lambda_{nk}\b{t-s}$ to first order in $s$ near the instantaneous time $t$ 
\begin{equation}
    \Lambda_{nk}\b{t-s}\approx \Lambda_{nk}\b{t} -\f{d\Lambda_{nk}\b t}{dt} s~~. 
    \label{eq:Lambda_approx}
\end{equation}
This approximation is justified if the bath dynamics are fast relative to the change in $\Lambda_{nk}\b t$, or alternatively, the change in the driving. In this regime, when $s$ is comparable, or greater than, the bath typical timescale, $s\sim\tau_B$, the contribution to the integral is negligible due to the fast decay of the bath correlation functions (Markovianity of the bath). In addition, from the inertial condition, the coefficients $\xi_{nk}$ and eigenoperators $\hat{F}_k$ depend weakly on time (only through $\v{\chi}\b t$), enabling the approximations $\xi\b{t-s}\approx \xi\b t$ and $\hat{F}_k\b{\v{\chi}\b{t-s}}\approx \hat{F}_k\b{\v{\chi}\b t}$.  These approximations are justified for Markovian dynamics.

We now perform the rotating wave approximation, which terminates non-conserving energy terms. Finally, in the spirit of Ref. \cite{breuer2002theory} and neglecting the Lamb-shift term, the non-adiabatic master equation becomes
 \begin{multline}
     \f d{dt}\tilde{\rho}_{S}\b t=
    \sum_k r_k\bigg( \hat{F}_{k}\b{\v{\chi}\b t} \tilde{\rho}_{S}\b t\hat{F}_{k}^{\dagger}\b{\v{\chi}\b t}\\-\{\hat{F}_{k}^{\dagger}\b{\v{\chi}\b t}\hat{F}_{k}\b{\v{\chi}\b t},\tilde{\rho}_{S}\b t\}\bigg) ~~,
    \label{eq:gen_NAME}
 \end{multline}
 with
 \begin{equation}
     r_k = \sum_{n,n'}\gamma_{nn'}
     \b{\Lambda_{n,k}'\b t}\xi_{nk}\b t\xi_{n'k}\b t~~,\\
     \nonumber
 \end{equation}
 where $\gamma_{nn'}\b{\alpha}=\infint ds\,e^{i\alpha s}\mean{B^\dagger_n\b s B_{n'}\b 0}$ is the  Fourier transformation of the bath correlation functions.
 The positivity of $\xi_{n,k}$ ensures that the NAME is of the GKLS form, guaranteeing a completely-positive trace-preserving map \cite{lindblad1976generators,gorini1976completely}. For a more detailed derivation, see Ref. \cite{dann2018time}.

\subsection{Validity regime of the NAME}
\label{subsec:validity_NAME}
The various approximations performed throughout the construction determine the validity regime of the final equation of motion. These approximations involve four timescales: (i) The system typical time scale $\tau_S\sim \omega^{-1}$, which is proportional to the inverse of the system's Bohr frequencies; (ii) the timescale characterizing the decay of the bath correlation functions $\tau_B$; (iii) the System relaxation timescale, which scales with the square of the system-bath coupling constant, $\tau_R\propto g^2$, in the weak coupling limit; (iv) the driving timescale, identified as $\tau_d=\text{min}_{n,k,t}\sb{\b{d\Lambda_{n,k}\b t/dt}/\b{d^2\Lambda_{n,k}\b t/dt^2}}$.

Typically, Master equations are valid under a coarse-graining of time, neglecting memory effects within the bath relaxation time. Thermodynamically, such an approach is related to the isothermal partition of system and bath, and is valid in the weak coupling regime and manifested by the Born-Markov approximation. Weak coupling between system and bath, $g\ll1$, implies a slow relaxation relative to both system and bath internal dynamics, $\tau_R\gg \tau_S,\tau_B$. Moreover, the Born-Markov approximation is valid when the bath dynamics is much faster than the system dynamics, implying $\tau_B\ll\tau_S$. Following these approximations, fast oscillating phases are terminated by the rotating wave approximation, which requires that $\tau_R\gg\tau_S$. Finally, the last assumption involves neglecting higher order terms in the expansion of $\Lambda\b{t-s}$ and $\xi\b{t-s}$. This assumption is justified when the bath correlation functions decay rapidly in comparison to the change in the driving, $\tau_B\ll\tau_d$. 

The hierarchy between the four time-scales determines the validity regime:
\begin{equation}
    \tau_R\gg\tau_B\,\,\,,\,\,\,\tau_S\gg\tau_B\,\,\,,\,\,\,\tau_R\gg\tau_S\,\,\,,\,\,\,\tau_d\gg \tau_B~~.
\end{equation}
This implies that the NAME is exact in the weak coupling limit and a  delta-correlated bath.

\subsection{Eigenoperators connection to the Lindblad jump operators}
\label{subsec:connection eigenoperator jump}

The eigenoperators of the free evolution operator play a key role in the description of open quantum system dynamics. For example, according to the Davis construction \cite{davies1974markovian} of the static Master equation, population changes are induced by the ladder operators connecting energy eigenstates $\ket{\eps_k}$ and $\ket{\eps_j}$. As the Hamiltonian is time-independent, the eigenstates of the free propagator and the Hamiltonian coincide and, as a result, the eigenoperators of the free propagator constitute the jump operators of the master equation. For a time-dependent Hamiltonian  under periodic driving, the eigenoperators of the free propagator are the ladder operators between the Floquet states \cite{alicki2012periodically}, which, according to a first principle derivation, constitute the jump operators of the Floquet Master equation \cite{alicki2012periodically}.

These examples lead us to interpret the eigenoperators of the free evolution under non-adiabatic driving as the ladder operators of the associated "time-global" eigenstates, which account for the integration and time-ordering procedure in the free propagator. In fact, under non-adiabatic driving, the bath induces transformations with respect to those time-global states, and therefore, it is natural that the eigenoperators of the free propagator constitute the jump operators of the NAME.

\subsection{Properties of non-adiabatic open system dynamics}
\label{subsec: properties of non-adiabatic open systems}
Non-adiabatic features appear in the kinetic coefficients and Lindblad jump operators. Both depend on the driving protocol, and differ from the equivalent terms in the adiabatic master equation \cite{albash2012quantum,dann2018time}. These terms define an instantaneous attractor $\hat{\rho}_{I.A}$
\begin{equation}
    {\cal{L}}\b t  \hat{\rho}_{I.A}\b t=0~~~.
\end{equation}
This attractor is a generalized Gibbs state beyond the adiabatic limit.
In general, any completely positive map has an invariant, which may not be unique \cite{lindblad1975completely}. Based on the Davis construction, we conjecture that for a non-degenerate Hamiltonian, the instantaneous attractor of the NAME is unique \cite{alicki2018introduction}.

 \section{Quantum control of open systems}
 \label{sec:control}
 Fast thermalization  can be formulated as a  quantum control problem  \cite{koch2016controlling}. 
 The task is classified as a state-to-state objective
 $\hat \rho_S^i \rightarrow \hat \rho_S^f$, governed by open system dynamics:
 \begin{equation}
     \frac{d}{dt} \hat \rho_S = {\cal L}\b t \hat \rho_S~~.
     \label{eq:drhodt}
 \end{equation}
 We consider a control problem where the reduced system dynamics is generated by the NAME (Eq. \eqref{eq:gen_NAME}).

Quantum control theory \cite{koch2016controlling,d2007introduction,glaser2015training} addresses three main topics:
 \begin{itemize}
 \item{Controllability, i.e, the conditions on the dynamics that allow obtaining the 
 state-to-state objective;} 
 \item{Constructive mechanisms of control, the problem of synthesis;}
 \item{Optimal control strategies and quantum speed limits.}
 \end{itemize}
 These topics will be employed to analyze the process of fast thermalization. 
 
 \subsection{State-to-state controllability}
 \label{subsec: state to state control}
 
 Under what conditions can an initial quantum state $\hat \rho_S^i$ be transformed into a final state $\hat \rho_S^f$, employing open systems dynamics?
To prove controlability a single explicit solution for the control task is sufficient, under the dynamics of Eq. \eqref{eq:drhodt}. In the context of the proof, the solution protocol has no time or energy restrictions.
 Entropy change is a necessary condition for  open system controlability. This requirement leads to a natural decomposition of the protocol: an entropy-changing dissipative part followed by a fast unitary conserving entropy:
 \begin{enumerate}
     \item For any initial state $\hat \rho_S^i$, couple the system to a bath of temperature $T_B$ and
change the Hamiltonian to $\hat H_S^f$ until it reaches an equilibrium thermal state $\hat \rho_{S,T_B}^f$, which has a common eigenvalue with $\hat{\rho}_S^f$;
\item Apply a fast unitary transformation $\hat U_S$, such that $\hat \rho_S^f = \hat U_S \hat \rho_{S,T_B}^f \hat U_S^{\dagger}$.
 \end{enumerate}
 In the first stage, a dissipative transformation that leads to common eigenvalues with $\hat{\rho}_S^f$ is obtained by reaching a thermal state $\hat \rho_{S,T_B}^f=\sum \lambda_j |\phi_j \rangle \langle \phi_j|$, determined by the Hamiltonian:
$\hat H_S^f = \sum_j \epsilon_j |\phi_j \rangle \langle \phi_j|$, where $\epsilon_j=kT (\log Z + \lambda_j)$. This state is unitarily equivalent to the target state, since unitary transformations do not alter the eigenvalues of the density matrix. 
 
 The protocol can be achieved under the following conditions:
 \begin{enumerate}
 \item{The isolated quantum system is completely unitary controllable, i.e., any unitary transformation $\hat U_S$ is admissible.}
 \item{There is a complete freedom in modifying the Hamiltonian $\hat H_S\b t$ of the system, embedded in the bath of temperature $T_B$.} 
 \end{enumerate}

The first controllability condition, open system entropy-changing control, can be addressed
in the framework of the GKLS dynamical equations (Eq. (\ref{eq:lindblad}).
The control of a target state $\hat \rho_S^f$ requires engineering 
the asymptotic  invariant of ${\cal L}$ to become unitarily equivalent to the target state \cite{lloyd2001engineering,lutkenhaus1998mimicking,murch2012cavity,puthumpally2017towards,fischer2019time}.
In the adiabatic limit under the Davis construction, the thermal state with the Hamiltonian $\hat{H}_S^f$ becomes the invariant of the GKLS equation \cite{davies1974markovian}.

The second condition, unitary control of a closed quantum system, is formulated employing a Lie algebra \cite{d2007introduction,huang1983controllability,jurdjevic1972control}. In this case,
the Hamiltonian of the system is separated into drift and control terms,
\begin{equation}
 \hat H_S(t) = \hat H_0 + \sum_j u_j(t) \hat H^j~~, 
 \label{eq:conthamil}
 \end{equation}
where $\hat H_0$ is the free system Hamiltonian, $u_j(t)$ are the control fields and $\hat H_j$ are control operators.
The system is unitary controllable provided that the Lie algebra, spanned by
the nested commutators of $\hat H_0 $ and $\hat H^j$, is full rank \cite{d2007introduction,huang1983controllability,jurdjevic1972control}. 
Under this condition, an arbitrary unitary propagator can be obtained. Such transformation necessarily preserves the eigenvalues of $\hat \rho_S$.

Controllability of open systems has previously been addressed assuming the dissipative generator ${\cal{L}}_D$ which is independent of the system Hamiltonian $\hat H_S$. This case has limited controllability \cite{dirr2009lie,koch2016controlling,mukherjee2013speeding}, since there is no control of the invariant of $\cal{L}$.
In the control community, the dependence of ${\cal{L}}$ on the Hamiltonian has been mostly overlooked (an exception is \cite{d2014control}). This dependence is required for a consistent thermodynamic description \cite{alicki2018introduction}. An exception arises for a singular bath (infinite temperature): then  ${\cal L}_D$ is independent of $\hat{H}_S$ \cite{gorini1976n}. In this case the generator becomes 
${\cal L}_D=-\gamma [\hat G,[\hat G, \bullet ]]$.
Then, if $\hat G$ belongs to the set of control operators \eqref{eq:conthamil}, control is limited to a wedge in the state space \cite{dirr2009lie}.
A similar result is obtained if ${\cal L}_D$ is unital \cite{o2012illustrating}.

\subsection{Constructive mechanisms of control}
\label{subsec:constructive mechanisms of contorol}

There are two limiting opposing mechanisms of control: adiabatic and  quench.
The adiabatic protocol interpolates between $\hat H_S^i$ and $\hat H_S^f$, implying the general form
$\hat H_S(t)=\lambda(t) \hat H_S^i+(1-\lambda(t))\hat H_S^f$, where $\lambda(t)$ is a slowly varying function of time.
In this limit, the state remains in the canonical state $\hat \rho_S\b t = \frac{1}{Z}\exp\b{-\beta \hat H_S\b t}$, and no coherence is generated during the control procedure. 
This procedure is reversible, $\Delta {\cal S}_U=0$, implying null dissipation of work  $W_{diss} = W-\Delta F_S =0$, where $\Delta F_S$ is the change in free energy and ${\cal S}_U$ is the entropy of the universe.
However, the optimal work cost, obtained in the adiabatic limit, is never practical since it requires an infinitely long execution time.

The other extreme mechanism is a quench: a sudden jump from the initial $\hat H_S^i$ to the target Hamiltonian $\hat H_S^f$, followed by a relaxation toward equilibrium  \cite{quan2010testing}. Whenever $\sb{\hat H_S^i,\hat H_S^f} \ne 0$ the protocol is irreversible. 
For such a case, the sudden quench generates significant coherence, requiring additional work 
that is eventually wasted by dissipation to the bath. 
This leads to an increased work cost, $W = \text{Tr} \{ \hat \rho_S^i (\hat H_S^f -\hat H_S^i)\}$, relative to the adiabatic work.
The timescale of the quench protocol is dictated by the bath relaxation rates. 

In the present study, we propose a fast control mechanism that serves as an intermediate between the adiabatic and sudden (quench) limits. In this protocol the state follows a generalized canonical form
$\hat \rho_S\b t = \exp\b{ \sum_j \lambda_j\b{t} \hat A_j }$, where $\lambda_j$ are time-dependent coefficients
and $\{ \hat A_j \}$ are members of a closed Lie algebra which includes $\hat H_S\b t$. Asymptotically, for long protocol duration, this protocol converges to the adiabatic one.
The fast protocol can achieve fast thermalization  at low entropic cost.
The basic idea, underlying the control scheme, is reverse-engineering the control Hamiltonian $\hat H_S\b t$ such that a specific trajectory, defined by the generalized canonical form, is maintained.

\subsection{Optimal control and quantum speed limit}

Optimal control theory aims to find an accurate transformation of the system to the target state, 
subject to constraints of finite resources, such as time and external power.
Optimal control algorithms have been applied to open system dynamics. 
Previously, in  all cases studied, the dissipator ${\cal L}_D$ was independent of the control 
Hamiltonian $\hat H_S\b t$ \cite{bartana1997laser,bartana2001laser,ohtsuki1999monotonically,morzhin2019maximization,mukherjee2013speeding}.
In the present study, we do not impose optimality; nevertheless, our fast thermalization protocol approxies the target.
For the existing protocols, we can compare the actual protocol speed to theoretical bounds that limit the change in purity $P\b t=\text{tr}\b{\hat{\rho}_S^2\b t}$.

The speed limit of the purity change is given by
\begin{equation}
   \left| \text{ln}\b{\f{P\b{t_f}}{P\b{t_i}}}
    \right|\leq 4\int_{t_i}^{t_f}\sum_k||r_k\b{t'}\hat{F}_k\b{t'}||^2_{sp}dt~~,
    \label{eq:speed limit}
\end{equation}
where $\hat{F}_k$ and $r_k$ are the Lindblad jump operators and rates, Eq. \eqref{eq:gen_NAME}, and $||
\cdot||_{sp}$ denotes the spectral norm.

\section{STE of a two-level system}
\label{sec:STE for TLS}
We demonstrate the construction of a shortcut to STE protocol for a two-level system embedded within a thermal bath. The general derivation follows the steps explicitly given in Sec. \ref{sec:driven dynamics}. The first crucial step is to solve the free dynamics of the system, so as to obtain the propagator ${\cal U}_S\b t$.

\subsection{Inertial dynamics of the isolated two-level system}
\label{subsec:Inertial_dynamics_TLS}
We consider a two-level-system (TLS), in the presence of two orthogonal modulated fields. The system is represented by the Hamiltonian 
\begin{equation}
    \hat{H}_S\b t= \omega\b t \hat{S}_z+ \eps \b t\hat{S}_x~~,
\label{eq:Ham_TLS}
\end{equation}
where $\hat{S}_i$, $i=x,y,z$ are the spin operators. Generally, the dynamics of a system with a time-dependent Hamiltonian is given by the time-ordered propagator, Eq. \eqref{eq:torder}.  However, this expression is only formal and is impractical for our analysis. We circumvent the time-ordering problem by utilizing the inertial theorem and solution \cite{dann2018inertial}.

The equations of motion for the $\mathfrak{su}$(2) algebra can be cast into the desired form, Eq. \eqref{eq:inertial_decompasition}, by choosing an appropriate time-dependent operators basis and protocol. To this end, we introduce two additional operators: $\hat{L}\b t=\eps\b t \hat{S}_{z}-\omega\b t \hat{S}_{x}$ and $\hat{C}\b t=\bar{\Omega} \b t \hat{S}_{y}$, where $\bar{\Omega}\b t=\sqrt{\omega^2\b t+\eps^2\b t}$ is the generalized Rabi frequency of the TLS,  and define the Liouville vector
\begin{equation}
\v v\b t=\{\hat{H}_S\b t,\hat{L}\b t,\hat{C}\b t\}^T~~~.
\label{eq:v_TLS}
\end{equation}
 Along with the identity $\hat{I}$, the elements of vector $\v v\b t$ form a basis of the Liouville space, completely determining the system dynamics. For a protocol satisfying a constant adiabatic parameter
\begin{equation}
\mu\equiv\f{\dot{\omega}\eps-\omega\dot{\eps}}{\bar{\Omega}^{3}}~~,
\label{eq:mu}
\end{equation}
 the dynamical equation in Liouville space is of the form $\f{d\v v}{dt} =\b{\f{\dot{\bar{\Omega}}}{\bar{\Omega}^{2}}{\cal{I}}+ \bar{\Omega} \b t {\cal{B}}\b\mu}\v{v}$, where ${\cal{I}}$ is the identity matrix in Liouville space. Expressing the dynamics in terms of the scaled vector, $\v w\b t  \equiv \f{\bar{\Omega}\b 0}{\bar{\Omega} \b {t}} \v v\b t$, leads to the required decomposition, Eq. \ref{eq:inertial_decompasition}: $\f{d\v{w}}{dt}={\cal{M}}\b{t}\v{w}$, with  ${\cal{M}}\b t =  \bar{\Omega} \b t {\cal{B}}\b\mu$. ${\cal{B}}\b{\mu}$ and the diagonalizing matrix $\cal{V}$ are given by
 \begin{equation}
     {\cal{B}}\b{\mu} = i\sb{\begin{array}{ccc}
0 & \mu & 0\\
-\mu & 0 & 1\\
0 & -1 & 0
\end{array}}~~\,\,\,,\,\,\,{\cal{V}}=\sb{\begin{array}{ccc}
1 & -\mu & -\mu\\
0 & i\kappa & -i\kappa\\
\mu & 1 & 1
\end{array}}~~.
 \end{equation}
Such a solution contains a single constant parameter, $\v \chi =\chi =\mu$. 

Next, we utilize the inertial theorem to obtain the dynamics of the eigenoperators of the free propagators ${\hat{F}_k=\{\hat{\xi},\hat{\sigma},\hat{\sigma}^{\dagger} \}}$
\begin{multline}
\hat{\xi}\b{\mu,0}= \f{1}{\kappa \bar{\Omega}\b 0}\b{\hat{H}_S\b 0+\mu \hat{C}\b 0} \\
\hat{\sigma}\b{\mu,0} = \f{1}{2\kappa^2\bar{\Omega}\b 0}\b{-\mu \hat{H}_S\b 0 -i \kappa \hat{L}\b 0+\hat{C}\b 0}
\label{eq:eigenoperators_TLS}
\end{multline}
and $\sigma^\dagger\b{\mu,0}$, with corresponding eigenvalues $\lam_1 = 0,\lam_2\b{t} = \kappa\b{\mu\b t}$ and $\lam_3\b{t} = -\kappa{\b{\mu{\b t}}}$, where
\begin{equation}
\kappa\b{\mu \b t}=\sqrt{1+\mu^2\b{t}}~~. 
\label{eq:kappa}
\end{equation}
For a slow change in $\mu$ the eigenoperators evolve according to the inertial solution $\v{v}\b t$, with vector elements
\begin{multline}
    \hat{v}_i\b{t}=
    \f{\bar{\Omega}\b t}{\bar{\Omega}\b 0}\sum_{j,k=1}^3{\cal{V}}_{ik}\b{\mu\b t}e^{-i\int_{0}^{ t}{\lam_k\b{t'}\b{\f{d \bar{\theta}\b{t'}}{dt'}}dt'}}\\
    \times{\cal{V}}^{-1}_{kj}\b{\mu \b t} \hat{v}_j\b{0}~~,
    \label{eq:inertial_TLS-1}
\end{multline}
with $\hat{F}_k\b{\mu\b t,0}={\cal{V}}^{-1}_{kj}\b{\mu \b t} \hat{v}_j\b{0}$ and $\bar{\theta}\b{t}=\int_0^t\bar{\Omega}\b{t'}dt'$.
 In the regime $\f{d\mu}{dt}\ll 2\kappa^2 \b{\f{d\bar{\theta}}{dt}}$, this solution describes the system dynamics accurately and is analogous to Eq. \eqref{eq:gen_inertial_theorem} of the general derivation.

To understand the practical implication of the inertial condition recall the role of the  adiabatic parameter $\mu$. This parameter naturally characterizes the qualitative accuracy of the adiabatic solution, evaluating the driving speed relative to the system Bohr frequencies. Hence, a slow change in $\mu$ corresponds to slow `acceleration' in the drive. Note that Eq. \eqref{eq:inertial_TLS-1} is also valid for fast driving (non-adiabatic), with large $\mu$, as along as the inertial condition is satisfied.

 \subsection{Inertial open system dynamics for the two-level system}
For the studied example, we consider a  system interacting with an Ohmic Boson bath at temperature $T_B$, through a single spin component
\begin{equation}
\hat{H}_I = g \hat{S_y}\otimes{\hat{B}}~~.
\label{eq:Ham_int_TLS}    
\end{equation}
Similarly, an alternative interaction can be assumed, leading to the same qualitative results. To derive the NAME we first perform a transformation to the interaction representation. This is achieved by noticing that $\hat{S}_y=\hat{C}\b t/\bar{\Omega}\b t=\hat{v}_3/\bar{\Omega}\b t$, and gathering Eqs. \eqref{eq:eigenoperators_TLS} and \eqref{eq:inertial_TLS-1}.
Following the derivation described in Sec. \ref{subsec:Open_system_dynamics}, the explicit expression for $\tilde{H}_I\b t$ leads a non-adiabatic Master equation which is analogous to Eq. \eqref{eq:gen_NAME}. The Master equation for the two-level system reads
\begin{multline}
    \f d{dt}\tilde{\rho}_{S}\b t=k_{\downarrow}\b t\b{\hat{\sigma}\b{\mu}\rho_{S}\hat{\sigma}^{\dagger}\b{\mu}-\f 12\{\hat{\sigma}^{\dagger}\b{\mu}\hat{\sigma},\tilde{\rho}_{S}\}}\\+k_{\uparrow}\b t\b{\hat{\sigma}^{\dagger}\b{\mu}\tilde{\rho}_{S}\hat{\sigma}-\f 12\{\hat{\sigma}\b{\mu}\hat{\sigma}^{\dagger}\b{\mu},\tilde{\rho}_{S}\}}~~,
    \label{eq:NAME_TLS}
\end{multline}
with kinetic coefficients
\begin{gather}
k_{\uparrow}\b t=\f{2\alpha\b t}{\hbar c}N\b{\alpha\b t}=k_{\downarrow}\b t e^{-\alpha\b t/k_B T}~~,
\label{eq:kinetic_coeff}\\
    \alpha\b t=\kappa\b t\bar{\Omega}\b t
    \label{eq:alpha}
\end{gather}
and $\hat{\sigma}\b{\mu}\equiv \hat{\sigma} \b{\mu\b t,0}$ depends on the instantaneous adiabatic parameter $\mu\b t$. 
Equation \eqref{eq:kinetic_coeff} has a similar structure to the detailed balance relation exhibited in the adiabatic master equation \cite{albash2012quantum}. However, beyond the adiabatic regime the transition frequency is increased by $\kappa\b t$, Eq. \eqref{eq:alpha} which is always greater than unity, Eq. \eqref{eq:kappa}. This property is a consequence of the influence of the external driving on the dissipation.

\subsection{Solution and properties of driven open systems}
\label{subsec:Evolution_and_properties} 
To solve the dynamics, we choose to parameterize the density operator as a generalized Gibbs state in a generalized canonical form
\begin{equation}
    \tilde{\rho}_S\b{t}=Z^{-1}e^{\gamma\b t \hat{\sigma}\b{\mu}}e^{\ss\b t \hat{\xi}\b{\mu}}e^{\gamma*\b t \hat{\sigma}^\dagger\b{\mu}}~~,
    \label{eq:rho_TLS}
\end{equation}
where $\tilde{Z}\equiv \tilde{Z}\b t=\text{tr}\b{\tilde{\rho}_S\b t}$  is the partition function, with time-dependent parameters $\gamma\b t$ and $\ss\b t$.
Since $\{\hat{\xi}\b{\mu},\hat{\sigma}\b{\mu}, \hat{\sigma}^\dagger\b{\mu},\hat{I} \}$ span the operator Hilbert space, Eq. \eqref{eq:rho_TLS} serves as a complete description of the TLS state. A linear parameterization is an additional option, leading to an alternative set of coupled equations \cite{kosloff2010optimal}.

Another common solution strategy is a formulation of the dynamics in the Heisenberg representation. However, when the Liouvillian ${\cal{L}}\b t$ (generator of the reduced open system dynamics) depends explicitly on time, as in Eq. \eqref{eq:NAME_TLS}, this framework requires a time ordering procedure \cite{breuer2002theory}.

To solve Eq. \eqref{eq:rho_TLS} we substitute Eq. \eqref{eq:rho_TLS} into the Eq.\eqref{eq:NAME_TLS} and obtain a set of coupled differential equations in terms of $\gamma\b t$  and $\ss\b t$
\begin{multline}
\dot\ss =\frac{\gamma e^{\ss}}{2\kappa^{2}}\dot{\gamma}^{*} -k_{\downarrow}\frac{4\kappa^{2}\left(e^{\ss}+1\right)+|\gamma|^{2}e^{\ss}}{16\kappa^{4}}\\+k_{\uparrow}\frac{\left(|\gamma|^{2}+e^{-\ss}4\kappa^{2}\right)\left(4\left(e^{\ss}+1\right)\kappa^{2}+e^{\ss}|\gamma|^{2}\right)}{64\kappa^{6}}
\label{eq:beta_comp}
\end{multline}
\begin{equation}
\dot{\gamma} = k_{\downarrow}\frac{\gamma}{8\kappa^{2}}-k_{\uparrow}\frac{\gamma\left(2\left(1+2e^{-\ss}\right)\kappa^{2}+|\gamma|^{2}\right)}{16\kappa^{4}}~~,
\label{eq:gamma_comp}
\end{equation}
where the $\ss$, $\gamma$, $k_\downarrow$ and $k_\uparrow$ are explicitly time-dependent and $\kappa\b t$ varies slowly with $\mu\b t$.
These equations are an alternative representation of the open system dynamics.

The evolution is given in the interaction picture relative to the free dynamics. In this picture, the system rotates with the isolated driven system, so any change to the density operator is induced solely by the interaction with the bath.

\subsection{Design of the control protocol}
\label{subsec:design_of_protocol}
The driving generated by the control has both a direct impact on the state through the unitary part as well as an indirect influence through the dissipative part, Sec. \ref{subsec:Open_system_dynamics} and \ref{sec:control}. To overcome this convoluted control scenario, we employ a reverse-engineering approach.

We consider the following control scenario. Initially, the system is in a Gibbs state, characterized by  $\gamma\b 0=0$ with Hamiltonian $\hat{H}_S\b  0=\hat{H}_S^i$ and temperature $T_i$. At the initial time, the system is coupled to a thermal bath of temperature $T_B$. We wish to construct a control protocol $\hat{H}_S\b{t}$ that drives the system to a Gibbs state at temperature $T_f$ with the target Hamiltonian $\hat{H}_S\b{t_f}=\hat{H}_S^f$. In general, $T_B$, $T_i$ and $T_f$ may differ from one another, that is, the initial and final states are not required to be in equilibrium with the bath, just in a Gibbs form. 
The assumption of an initial Gibbs state simplifies the analysis but does not modify the system's behaviour qualitatively.

Since  Eq. \eqref{eq:gamma_comp} vanishes when $\gamma \b 0 =0$, the state conserves the form 
\begin{equation}
\tilde{\rho}_S\b t=\tilde{Z}^{-1}e^{\ss\b t \hat{\xi}\b{\mu}}
\label{eq:rho_beta}
\end{equation}
throughout its evolution, and the equation of motion is reduced to a single differential equation 
\begin{equation}
     \dot{\ss} \b t=  \frac{1}{4\kappa^{2}\b{\mu\b t}}\sb{k_{\uparrow}\b t\b{1+e^{-\ss\b t}}-k_{\downarrow}\b t\left(e^{\ss\b t}+1\right)}~.
     \label{eq:beta}
\end{equation}
This equation is the basis for the suggested control scheme.

The control strategy is as follows. We perform a change of variables $y\b t=e^{\ss{\b t}}$, and introduce a polynomial solution for $y\b t$ which satisfies the boundary conditions of the control. The initial and target Gibbs states correspond to values $\ss\b 0= - \hbar \bar{\Omega}_i/k_B T_i$ and $\ss\b{t_f}= -\hbar \bar{\Omega}_f/k_B T_f$, where $\bar{\Omega}_i=\bar{\Omega\b 0}$ and $\bar{\Omega}_f=\bar{\Omega}\b{t_f}$ are the generalized Rabi frequencies of the initial and final Hamiltonians, respectively. In addition, the values of $\dot{\ss}\b 0$ and $\dot{\ss}\b{t_f}$ are set by Eq. \eqref{eq:beta} and the requirement of stationary control at the initial and final times, $\mu\b 0 =\mu\b{t_f}=0$. Since the stationary condition implies $\dot{\alpha}\b 0 = \dot{\alpha}\b{t_f}=0$, the second derivatives $\ddot \ss \b 0$ and $\ddot \ss \b{t_f}$ are determined as well through the kinetic coefficients $k_\uparrow\b{\alpha{\b t}}$, $k_\downarrow\b{\alpha{\b t}}$, respectively (Eq. \eqref{eq:kinetic_coeff}).

A fifth-order polynomial for $y$ is sufficient to satisfy the boundary condition of Eq. \eqref{eq:beta}, leading to
\begin{equation}
    \ss \b t = {\text{ln}\b{\sum_{k=0}^5 b_k t^{k}}}~~,
    \label{eq:beta_solution}
\end{equation}
where the coefficients $\{b_k\}$ are presented in Appendix \ref{apsec:control_protocol}.
Next, we substitute the solution, Eq. \eqref{eq:beta_solution}, and the kinetic coefficients Eq. \eqref{eq:kinetic_coeff}  into Eq. \eqref{eq:beta}, which leads to an equation in terms of $\alpha\b t$.  We solve for $\alpha\b t$ using a standard numerical solver. Finally, by numerically reversing Eq. \eqref{eq:alpha}, we obtain the control protocol $\bar{\Omega}\b {\alpha \b t}$.

To gain some insight about the dynamics, we compute the instantaneous attractor of Eq. \eqref{eq:beta}. The attractor $\ss_{I.A}\b t$ is defined as the value of $\ss\b t$ for which $\dot{\ss}\b t = 0$ (simply  calculated by setting the LHS of Eq. \eqref{eq:beta} to zero). When $\ss \b t=\ss_{I.A}\b t$, the system is stationary. As the dynamical map is a semi-group, the  contraction property  \cite{goldstein1981one,pazy2012semigroups}, implies that at each instant, the system propagates towards the instantaneous attractor state $\tilde{\rho}_S\b{\ss_{I.A}\b t}$. 

For sufficiently slow driving, $\mu\ll 1$ (adiabatic regime), $\hat{\xi}\b{\mu \b t}$ converges to the instantaneous Hamiltonian $\hat{H}_S\b 0/\bar{\Omega} \b 0 $,  and Eq. \eqref{eq:beta} describes a decay towards the adiabatic attractor, $\ss_{I.A}\b t\ra -\hbar \bar{\Omega} \b t / k_B T_B$. This means that at long times, the system state converges to the adiabatic result  $\tilde{\rho}_S\b t =\exp\b{ -\hbar \bar{\Omega} \b t \hat{H}_S\b 0/ \sb{\bar{\Omega} \b 0 k_B T_B}}$.
Moreover, in the adiabatic limit, the relaxation rate towards the instantaneous thermal state reduces to the adiabatic rate, $k =k_{\downarrow}^{adi}\b t+k_{\uparrow}^{adi}$, where  $k_{\downarrow,\uparrow}^{adi}=k_{\downarrow,\uparrow}\b{\bar{\Omega} \b t}$, as $\alpha\b t\ra \bar{\Omega} \b t$ (see Eq. \eqref{eq:kinetic_coeff}).

Conversely, beyond the adiabatic limit, $\mu\cancel{\ll}1 $, $\hat{\xi}\b{\mu \b t}$ is a linear combination of $\hat{H}_S\b 0$ and $\hat{C}\b 0$ (see Eq. \eqref{eq:eigenoperators_TLS}). As a result, the instantaneous attractor state is a rotated squeezed Gibbs state that differs from the instantaneous thermal state. Thus, due to the non-adiabatic driving, the map propagates the system toward a state that differs from equilibrium. Overall, this behaviour can be understood as a dressing of the system by the drive, and consequently the bath interacts with a dressed system and leads the system toward the instantaneous attractor. 

In practice, when $\mu\sim 1$, $\ss\b t$ does not vary sufficiently rapidly so to follow $\ss_{I.A}\b t$.
As a result, the system in the interaction picture remains  a rotated squeezed Gibbs state with varying temperature throughout its evolution. 

The relaxation rates are also altered by the non-adiabatic driving. Beyond the adiabatic regime they do not depend on the instantaneous generalized Rabi frequency, but on the effective frequency $\alpha \b t$ (Eq. \eqref{eq:kinetic_coeff}). This leads to a skewed detailed balance and modifies the dissipation of energy and coherence. 

\paragraph{Control protocol}
The control protocol is defined in terms of the generalized Rabi frequency $\bar{\Omega} \b t$, leaving ambiguity concerning the values of ${\omega}\b t$ and $\eps \b t$ (Eq. \eqref{eq:Ham_TLS}). For the current model, we choose
\begin{equation}
  \omega\b t=\bar{\Omega}\b t \cos\b{\Phi\b t}\,\,\,\,\,\text{and}\,\,\,\,\,
\eps\b t=\bar{\Omega}\b t \sin\b{\Phi\b t}~~,
\label{eq:omega_eps}
\end{equation}
where $\Phi$ is taken as a simple polynomial. Substituting Eq. \eqref{eq:omega_eps} into $\mu$, Eq. \eqref{eq:mu}, leads to the relation $\dot{\Phi}=-\mu\b t \bar{\Omega}\b t$, which determines a solution of the form  $\Phi\b t=a\b{t^2+b t^{3}}$, satisfying the stationary condition at initial and final times. Here, $b=-2/3 t_f$ and $a$ is a free parameter which should be sufficiently small to comply with the inertial condition $d\mu/dt \ll 2\kappa^2 \b{d\theta/dt}$. For the modeling, it is chosen as $a=10/t_f^2$, leading to $\mu\ra0$ for $t_f\ra0$.

\section{Results and Discussion}
\label{sec:Results and Discussion}
\subsection{Control}

The control scheme allows addressing various control tasks:
\begin{enumerate}
    \item  Transformation between two equilibrium states with different Hamiltonians (STE);
    \item Transformation between an initial non-equilibrium state to an equilibrium state, accompanied by a change in the Hamiltonian (STE);
    \item Transformation to a final state which is colder than the bath, $T_f<T_B$.
\end{enumerate}

\begin{figure}[htb!]
\centering
\hspace{-1.2cm}
\includegraphics[width=8cm]{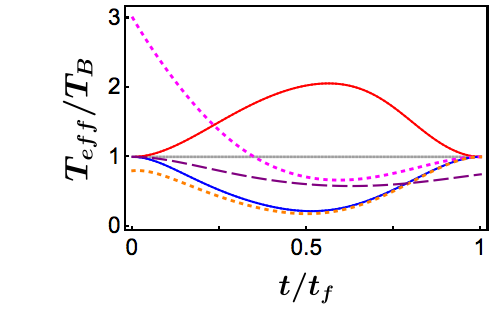}\\
\captionsetup{justification=raggedright,singlelinecheck=false}
\caption{Effective temperature as a function of time for protocol duration $t_f=6\,\b{2\pi/\bar{\Omega}_{ref}}$, for various STE protocols. The protocols are designated as follows: PE (blue) - expansion with initial and final thermal states ($T_i=T_f=T=5$); PC (red) - compression with initial and final thermal states; PE1 (dashed pink) expansion initializing at a non-equilibrium Gibbs state, with temperature $T_i=15$, with a final equilibrium state; PE2 (dashed orange) expansion of an initial Gibbs state with $T_i=4$ and a thermal target state $T_f=T=5$; PEC (long-dashed purple) - expansion initializing at a thermal state with a final state colder than the bath, $T_f=4$ (temperatures are given in atomic units). 
The gray continuous line designates $T_{eff}=T_B$, given as a reference.}
\label{fig:T_eff_vs_t}
\end{figure}

We study two classes of STE protocols, expansion and compression, with bath temperature $T_B=5$. 
The  expansion varies the generalized Rabi frequency from $\bar{\Omega}_i=12$ to $\bar{\Omega}_f=5$  (atomic units). Compression involves an increase in the generalized Rabi frequency: $\bar{\Omega}_i=5\ra\bar{\Omega}_f=12$.
During the expansion (compression) protocols, the effective temperature $T_{eff}\equiv-\bar{\Omega}\b t/\ss\b t$ decreases below (increases above) the bath temperature at transient times. At the final stage of the protocol, the effective temperature returns back to the bath temperature and we obtain $T_{eff}=T_B$ at the final time, Fig. \ref{fig:T_eff_vs_t}. 

\begin{figure}[htb!]
\centering
\hspace{-1.2cm}
\includegraphics[width=8cm]{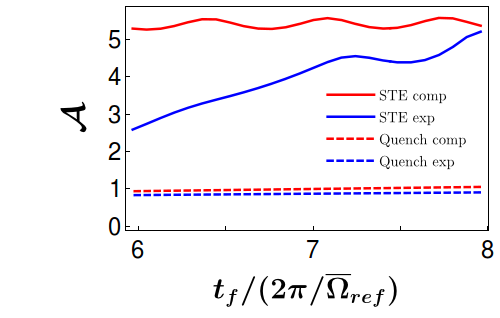}\\
\caption{Accuracy with respect to the control target as a function of protocol duration. The STE results correspond to expansion and compression protocols (PE and PC, respectively). These are compared to the analogous quench procedures. The accuracy is defined as ${\cal{A}}=-\log_{10}\b{1-{\cal{F}}\b{\hat{\rho}_S\b{t_f},\hat{\rho}_{S,T}}}$, and increases  with the fidelity, given by ${\cal{F}}\b{\hat{\rho}_1,\hat{\rho}_2}=\sb{\text{tr}\b{\sqrt{\sqrt{\hat{\rho}_1}\hat{\rho}_2\sqrt{\hat{\rho}_1}}}}^2$ \cite{uhlmann1976transition}. In the calculation, $\hat\rho_S\b{t_f}$ accounts for the deviations from the inertial approximation. The reference Rabi frequency is given by $\bar{\Omega}_{ref} = 5\,\text{a.u.}.$ }
\label{fig:accuracy_vs_tau_p}
\end{figure}

The shortcut to equilibrium protocols achieve high fidelity relative to the quench protocols for both expansion and compression procedures. Figure \ref{fig:accuracy_vs_tau_p} shows the accuracy $\cal A$ of the target state for varying protocol duration. Within the inertial approximation the control is precise and the fidelity $\cal F$ approaches unity. As a result, the inertial approximation is the limiting factor of the protocol accuracy. The fidelity is therefore evaluated by comparing the inertial solution for an isolated solution, under the shortcut protocol, to an exact numerical solution.

As the protocol duration increases, the inertial approximation improves and the accuracy increases. In comparison, the quench protocol leads to slow relaxation toward equilibrium, Sec. \ref{subsec:constructive mechanisms of contorol}. Shortcut protocols show an improvement of up to $5$-fold in accuracy, relative to the quench procedure for the same protocol duration.

\begin{figure}[htb!]
\centering
\includegraphics[width=6cm]{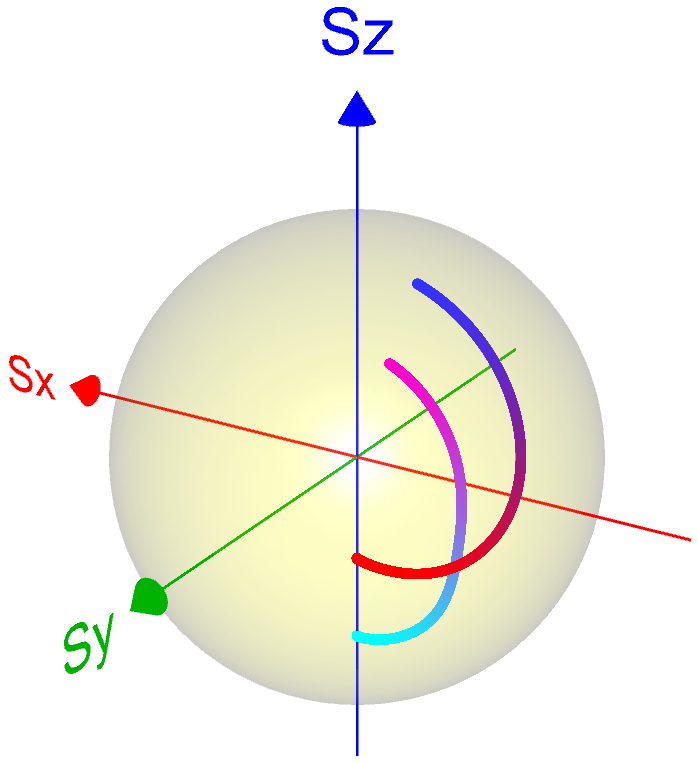}\\
\caption{Two-levels system trajectory in the spin operator space $\{\mean{\hat{S}_x},\mean{\hat{S}_y},\mean{\hat{S}_z}\}$. The ball represents a purity of 1. The compression trajectory (PC), red to blue, increases the purity and the expansion trajectory (PE), cyan to purple, decreases it.}
\label{fig:3D_S}
\end{figure}

The state trajectory in the $\{
\hat{S}_x,\hat{S}_y,\hat{S}_z\}$ space (Fig. \ref{fig:3D_S}) displays the change in purity imposed by the control protocol: an increase of purity for compression and a decrease for expansion.
The change in purity can be related to the speed limit (Eq. \eqref{eq:speed limit}); the bound on the purity change, for expansion and compression, is $1.58$, while the STE protocols yield a value of $0.34$. This value shows that although the STE is not optimum, its nevertheless within the range of the speed limit. 

\begin{figure}[htb!]
\centering
\includegraphics[width=7.5cm]{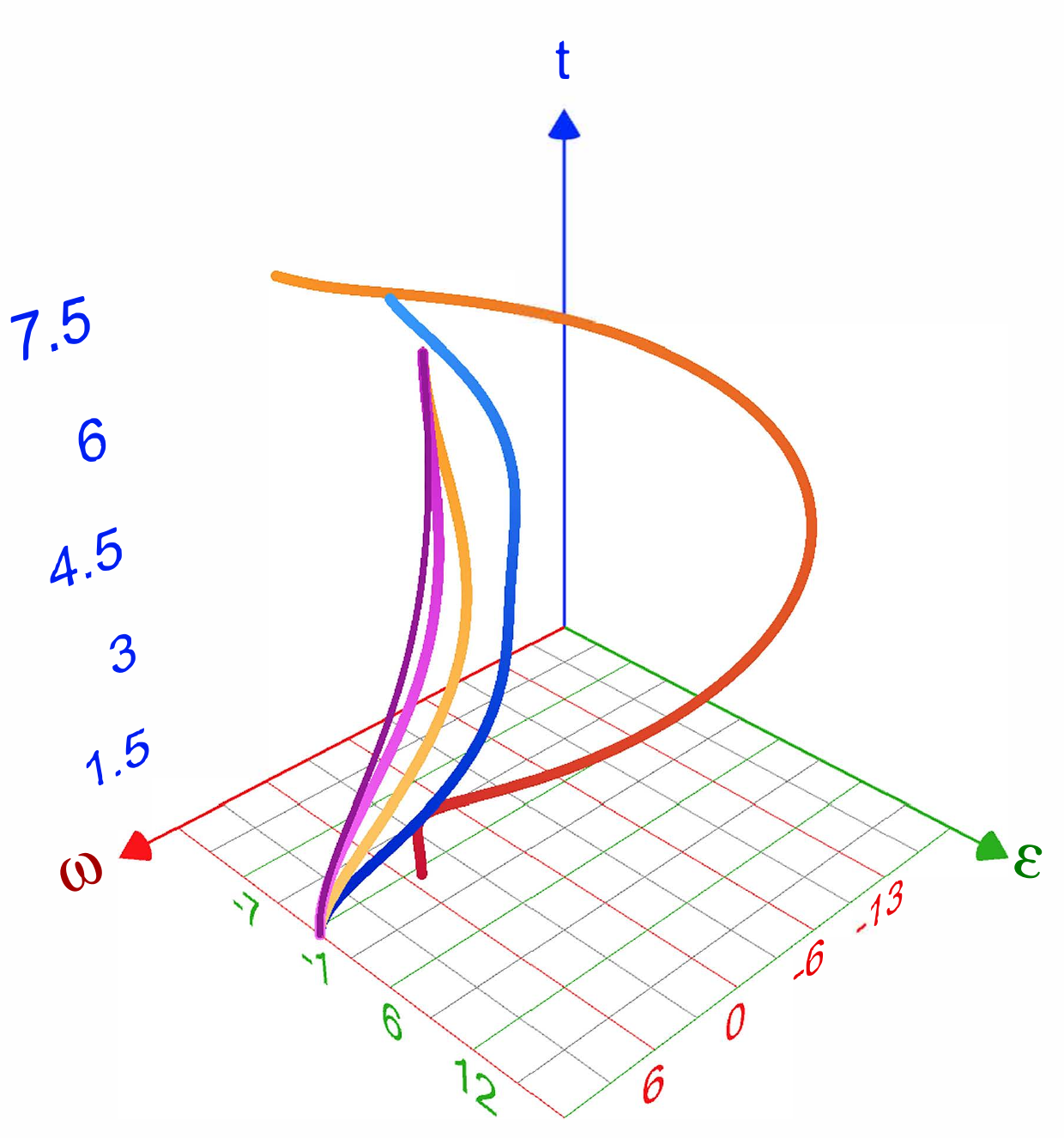}\\
\vspace{0.5cm}
\hspace{-1cm}
\includegraphics[width=9.5cm]{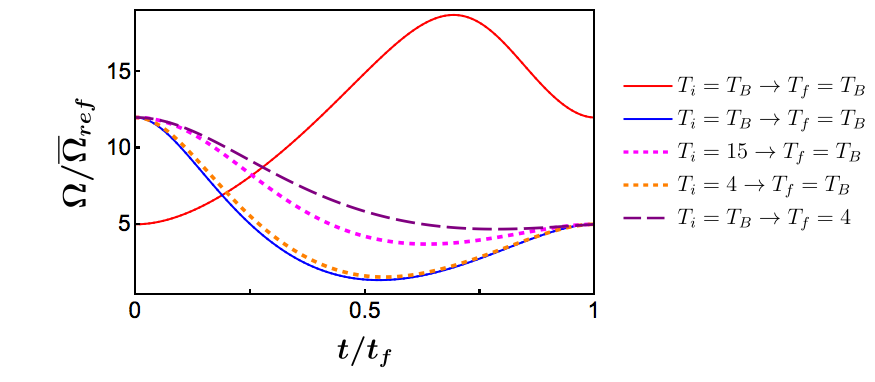}\\
\caption{Top: frequencies $\omega(t)$ and $\eps\b t$ as a function of time for expansion and compression protocols. The expansion protocols initialize at $\omega=12\, \text{a.u.}$ and $\eps =0 \, \text{a.u.}$, while the compression starts at $\omega =5 \, \text{a.u.}$ with $\eps=0 \, \text{a.u.}$. The various protocols differ by their initial temperature. Bottom: generalized Rabi frequency as a function of time for protocol duration $t_f=6\,\b{2\pi/\bar{\Omega}_{ref}}$, where $\bar{\Omega}_{ref}\equiv 5\,\text{a.u.}$. Both expansion (continuous blue, PE) and compression (continuous red, PC) protocols show overshoot with respect to the final Rabi frequency. The STE protocols and color code correspond to Fig. \ref{fig:T_eff_vs_t}.}
\label{fig:Omega_vs_time}
\end{figure}

The dynamics are imposed by non-trivial protocols (Fig. \ref{fig:Omega_vs_time}) of $\omega\b t$ and $\eps\b t$. These time-dependent frequencies are a single option from a family of procedures that correspond to the protocol $\bar{\Omega} \b t$ (see Sec. \ref{subsec:design_of_protocol}).

We find that the different protocols of compression and expansion for various initial states (Gibbs states with different temperatures) are all characterized by overshoot relative to the target frequency. As seen in Fig. \ref{fig:Omega_vs_time}, the expansion protocol (reduction in the generalized Rabi frequency $\bar{\Omega}\b t$) achieves values above the target frequency, whereas the compression first increases $\bar{\Omega}$, before a fast decline to $\bar{\Omega}\b{t_f}$. A similar behavior was witnessed for the STE protocol for a Harmonic oscillator \cite{dann2019shortcut}. Formally, this behavior is connected to the 
proposed ansatz, Eq. \eqref{eq:beta_solution}, and a different ansatz may alter this result. Nevertheless, this outcome fits the need to generate coherence, which in turn allows controlling the energy via a unitary transformation.

The STE protocol can be employed for an initial non-equilibrium Gibbs state.  For example, expansion protocols PE1 and PE2 begin in a Gibbs state with $T_i\neq T_B$ (see Figs.  \ref{fig:T_eff_vs_t} and \ref{fig:Omega_vs_time}). These STE protocols induce an energy exchange with the bath that leads the state toward an equilibrium state, with $T_{eff}\b{t_f}=T_B$. The fidelity of protocols PE1 and PE2 are comparable to those of PE and PC (see Fig. \ref{fig:T_eff_vs_t} legend and Table \ref{table:protocols} in the Appendix). 

Cooling is demonstrated in Fig. \ref{fig:T_eff_vs_t}, where the system reaches a final temperature of $T_f = 4\, \text{a.u.}$, below the bath temperature $T_B$. The final state is unstable once the driving ceases, that is, if the system remains in contact with the bath, it will equilibrate. The effect is shown in Fig. \ref{fig:entropy_production_vs_t}. We find that this control task is sensitive to the protocol, and further analysis is required to map the accessible cooling regime.

\subsection{Thermodynamic analysis}
\label{subsec:thermo analysis results}
The manipulation of a system Hamiltonian by external fields has an associated work cost or gain (gain or extracted work is considered a negative work cost).  This cost can be connected to quantum friction \cite{feldmann2003quantum,plastina2014irreversible}, which implies that faster transformations are accompanied by a higher energy cost \cite{chen2010transient,hoffmann2011time,salamon2009maximum,campbell2017trade,stefanatos2017minimum}.  We identify this cost with the microscopic work, defined as the time integral over the instantaneous power ${\cal{P}}\b t\equiv\mean{\pd{\hat{H}_S\b t}{t}}$. This interpretation is motivated by the correspondence between work and energy for an isolated system. In the absence of environmental interactions, the change in the system's energy can be formulated as $\Delta E_S\b{t}= \int_0^t {\cal{P}}\b{t'}dt'=W$, utilizing the Heisenberg picture for an isolated system. 
An extended definition also accounts for the cost of the external controller \cite{tobalina2019vanishing}. This additional cost requires an explicit description of the controller, which is beyond the scope of this study.

From conservation of energy, heat is then determined as $Q=\Delta E_S-W$, with $\Delta E_S=\text{tr}\b{\hat{H}_S\b{t_f} \hat{\rho}_S\b{t_f}}-\text{tr}\b{\hat{H}_S\b 0 \hat{\rho}_S\b 0}$. Utilizing the linearity of integration and the trace, this relation becomes $Q=\int_0^t {\cal{J}}\b{t'}dt'$, where  ${\cal{J}}\b{t'}\equiv\text{tr}\b{\hat{H}_S\b{t'}\f{d}{dt
'}\hat{\rho}_S\b{t'}}$ is the heat current.

\begin{figure}[htb!]
\centering
\hspace{-1.2cm}
\includegraphics[width=8cm]{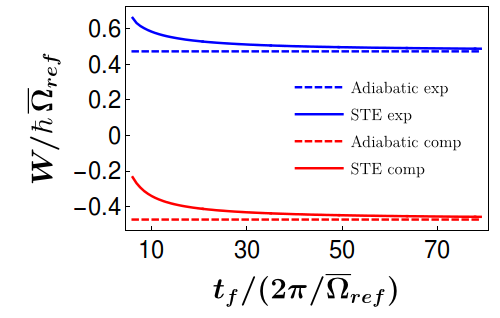}\\
\caption{Work as a function of the protocol duration $t_f$ for  expansion (blue, PE) and compression (red, PC) STE  procedures. Optimal values, obtained in the adiabatic limit ($t_f\ra \infty$), appear as dashed lines. Equilibration acceleration is accompanied by a work cost relative to the adiabatic result. This leads to an increased work cost for compression protocols, and reduced work extraction during expansion . }
\label{fig:work_vs_protocol_duration}
\end{figure}

\begin{figure}[htb!]
\centering
\hspace{-1.2cm}
\includegraphics[width=8cm]{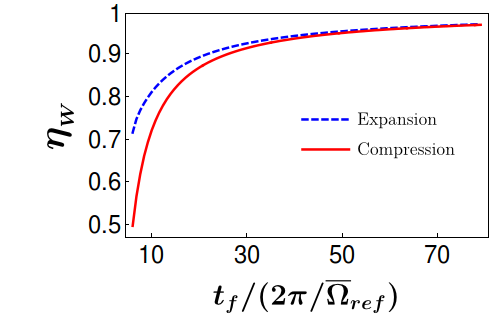}\\
\caption{Efficiency as a function of the protocol duration $t_f$ for  expansion (blue, PE) and compression (red, PC) STE  procedures. Color codes are defined in Fig. \ref{fig:T_eff_vs_t}. }
\label{fig:efficiency_vs_protocol_duration}
\end{figure}

The shortcut procedures accelerate the thermalization rate by investing additional work. Compared to the adiabatic protocol, the STE harvests less work in extraction (compression for the TLS) and requires an additional work cost under expansion (Fig. \ref{fig:work_vs_protocol_duration}) \footnote{Unlike typical working mediums, work is extracted from the TLS in a compression process and invested under expansion.}. The additional cost can be traced back to the transformation of invested work to coherence, which is then dissipated to the bath. As the protocol duration increases coherence generation is suppressed and the STE work cost converges to the adiabatic result.
To evaluate the relative performance, we define the work efficiency $\eta_W$, for an expansion of the two-level system $\eta_W=W_{adi}/W$, and for compression $\eta_W = W/W_{adi}$, where $W_{adi}$ is the adiabatic work. For fast protocols $t_f\sim 6\,\b{2\pi/\bar{\Omega}_{ref}}$, the work efficiency gives values of $\eta_W=0.71$ for expansion and $\eta_W=0.5$ for compression (see  Fig \ref{fig:efficiency_vs_protocol_duration}). For larger protocol duration, the work efficiency improves in accordance with the work.

\begin{figure}[htb!]
\centering
\hspace{-1.2cm}
\includegraphics[width=8cm]{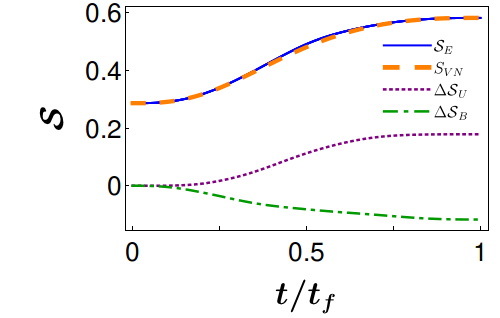}\\
\caption{Entropy as a function of time for an expansion procedure (PE) with protocol duration $t_f=6\,\b{2\pi/\bar{\Omega}_{ref}}$.  $\Delta {\cal S}_B$ is the change in the entropy of the bath, due to heat transfer. ${\cal S}_{VN}$ is the von-Neumann entropy of the system; the change in the entropy of the universe is $\Delta {\cal S}_U$ and ${\cal S}_E$ is  the system's energy entropy. ${\cal S}_E$ always exceeds the von-Neumann entropy as it does not contain information about the quantum correlations between the energy states.}
\label{fig:entropy}
\end{figure}

When the open quantum system, initially at thermal equilibrium, expands heat flows from the bath to the system, increasing its entropy. Figure \ref{fig:entropy} shows the increase in the system's von-Neumann entropy ${\cal{S}}_{V.N}$ along the expansion protocol, accompanied by a reduction in the bath entropy ${\cal{S}}_{B}=-{\cal Q}/T_B$. Their sum gives the total change in the entropy of the universe ${\cal{S}}_{U}$, which is strictly positive for any irreversible process. As expected, the fast driving during the STE protocol induces irreversible dynamics that increases ${\cal S}_{U}$. 

To evaluate the coherence generation, we compare ${\cal S}_{V.N}$ and energy entropy ${\cal S}_E=-k_B\sum_i p_i \text{ln}p_i $, where $\{p_i\}$ are the populations in the energy representation. As seen in Fig. \ref{fig:entropy}, the two are very close, demonstrating that for protocol duration $t_f= 6\,\b{2\pi/\bar{\Omega}_{ref}}$, the dynamics is dominated by the energy. Nevertheless, the two entropies differ at intermediate times when the state exhibits maximal coherence. At the beginning and final times ${ \cal S}_{V.N}={\cal S}_E$, since the state is diagonal in the energy representation.
\begin{figure}[htb!]
\centering
\hspace{-1.2cm}
\includegraphics[width=8cm]{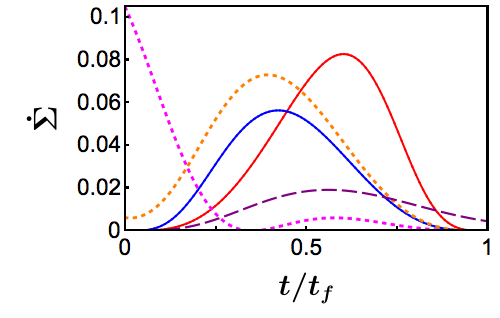}\\
\caption{Entropy production rate as a function of time for protocol duration $t_f=6\,\b{2\pi/\bar{\Omega}_{ref}}$. Red and blue curves correspond to expansion and compression procedures, respectively, with initial and final thermal states. When the system is in equilibrium with the bath, the entropy production vanishes ($t=0$ and $t=t_f$ for these protocols). Notice that the cooling protocol (long dashed-purple line) has a final positive entropy production rate as the system is in a non-equilibrium Gibbs state ($T_f=4\neq T_B$). The orange and pink protocols are associated with initial non-equilibrium  Gibbs states; orange and pink dashed lines correspond $T_i = 4$ and $T_f = 15$ (atomic units), respectively.}
\label{fig:entropy_production_vs_t}
\end{figure}

Another measure of the irreversibility is the entropy production $\Sigma$. Close to equilibrium, it can be defined as $\Sigma = {\cal S}_{V.N}-{\cal Q}/T_B$ or equivalently, as $W-\Delta F_S$. For the general case, quantum entropy production rate is recognized with the negative derivative of the  Kullback-Leibler divergence \cite{breuer2002theory} between the system state and the fixed point of the dynamical map (the instantaneous attractor). The divergence is a measure of the difference between two quantum states ${\cal D}\b{\hat{\rho}_1||\hat{\rho}_2}=\text{tr}\b{\hat{\rho}_1 \b{\text{ln}\hat{\rho}_1-\text{ln}\hat{\rho}_2}}$. It is non-negative and it decreases as the system evolves under Eq. \eqref{eq:lindblad} \cite{lindblad1975completely}. This property implies the positivity of the entropy production rate $ \dot{\Sigma}\b{\hat{\rho}_S\b t}\equiv-\f{d}{dt}{\cal D}\b{\hat{\rho}_S\b t||\hat{\rho}_S^{I.A}}\geq 0$. Figure \ref{fig:entropy_production_vs_t} presents the entropy production rate during a STE expansion protocol. As expected the entropy production rate is positive. It reaches maximum values at intermediate times, when coherence builds up and dissipates to the bath. 
This dissipation leads to the increase in total entropy and irreversibility. 
When the initial or final states are not in equilibrium with the bath (procedures PE1, PE2 and PEC), the entropy production does vanish at corresponding boundary.

The cooling mechanism  is analogous to that of a power driven refrigerator, where the two-level system mimics a cold bath with finite heat capacity. The driving supplies the power to pump heat from the cold bath and dump it into the hot bath. In such a cooling scenario, consistency with thermodynamics requires a positive entropy production and investment of work. This is verified in Fig. \ref{fig:entropy_production_vs_t}.

\section{Conclusions}
\label{sec:Conclusions}
Thermalization is typically considered a spontaneous process, the rate of which is determined by the system-bath coupling strength. We demonstrated that thermalization can be actively controlled, resulting in acceleration of the thermalization rate. 
Our control is achieved by implementing external driving that modifies the system Hamiltonian directly. First principle treatment (Sec. \ref{sec:driven dynamics}) showed that the  time-dependence of the system Hamiltonian dresses the system-bath interaction, modifying the dissipative part indirectly. This structure is vital for a dynamical description that complies with  thermodynamic principles \cite{alicki2018introduction}.  
The dissipative part adjusts to the change in the Hamiltonian and is therefore indirectly controlled.

Alternatively, direct control over dissipation can be obtained in the 'singular bath limit', i.e., an additional delta-correlated noisy driving with a fast timescale with respect to all other considered timescales \cite{schirmer2018robustness}. In this case the dissipation is independent of the system Hamiltonian. A recent study showed that a Langevin-type term, rising from momentum kicks, can lead to acceleration of the thermalization rate \cite{dupays2019shortcuts}. Another suggestion, is to include a counter-diabatic term to engineer the Hamiltonian and dissipative part \cite{alipour2019shortcuts}.

In a more generalized context, we incorporate the open-system control task into the theme of quantum control theory, Sec. \ref{sec:control}. Adopting the principles of the themalization control scheme, the analysis led to the formal conditions of complete state-to-state controllablity of open systems (Sec. \ref{subsec: state to state control}).

Speeding up the  themalization rate comes with a thermodynamic cost, with a  minimum work cost achieved in the adiabatic limit. Any additional cost is associated with the emergence of friction. When the Hamiltonian does not commute with itself at different times, driving at non-vanishing speed generates coherence and is manifested by additional work which is dissipated to the bath.  Dissipated energy heats the bath and leads to positive entropy production and irreversibility.  The first principle analysis conducted here demonstrated the quantum origin of friction.

To demonstrate the control scheme, we utilized a two-level-system model, characterized by an $\mathfrak{su}$(2) algebra.  Similarly, the analysis can be straightforwardly generalized to any system described by the same algebra and a Hamiltonian of the form of Eq. \eqref{eq:Ham_TLS}, such as, the four level system described in Refs. \cite{feldmann2003quantum,kosloff2002discrete,turkpencce2019coupled,peng2018influences}.
The present study originated from our previous analysis of the shortcut to equilibration (STE) of a harmonic oscillator \cite{dann2019shortcut}.  These control schemes have common properties: in both models, the fast protocols are characterized by overshoot with respect to the final control frequency. Furthermore, the thermodynamic cost shows a similar trend. In this study, we explored additional control protocols, extending the applicability of the control scheme. We showed that the control allows for non-equilibrium initial Gibbs states. Moreover, adding additional fast unitary transformations at initial and final times, allowed extending the control to a broad family of states containing the same initial and final entropies. 
Another surprising result is a protocol that leads to a final state that is colder than the bath ($T_f<T_B$). At first glance, this seems like a violation of thermodynamic principles, as it cannot be achieved under adiabatic driving.
However, the fast driving generates conditions analogous to a power-driven refrigerator, where heat is pumped from the two-level-system to the thermal bath by consuming external power. The mechanism requires generation of coherence whose dissipation to the bath generates entropy, the latter compensating for the negative entropy change of the system. We emphasize that this cold state is transient.

Experimental realization of quantum heat engines has emerged recently \cite{rossnagel2016single}. The platform employed in these experiments includes various controlled quantum systems. The protocols introduced above can be directly employed in the current devices to realize quantum heat engines at finite times. In particular, these protocols can be used to realize a quantum Carnot engine with a qubit as a working medium \cite{von2019spin}.

A natural extension of the present study; would involve incorporating optimal control theory in the control of an open quantum system.  The novelty of such an approach is the incorporation of control on the dissipation via external driving.

 \begin{acknowledgments}
 We thank KITP for their hospitality, this research was supported by the Adams Fellowship  Program of the Israel Academy of Sciences and Humanities and the Israel Science Foundation, grant number 2244/14, the National Science Foundation under Grant No. NSF PHY-1748958, the Basque Government, Grant No. IT986- 16 and  MINECO/FEDER,UE, Grant No. FIS2015-67161-P. 
 \end{acknowledgments}

\appendix
\section{Control protocol}
\label{apsec:control_protocol}
\break

\begin{table}
\caption{Model parameters.\label{table:model_parameters}}
\begin{ruledtabular}
\begin{tabular}{cccccccc}
 Coefficient& Value $\sb{\text{atomic units}}$ 
\\
\hline
Bath temperature & $5$ \\ 
Coupling prefactor &  $g^2/{2 \hbar c}=0.02$    \\
Numerical integrating step & $10^{-3}$  \\ 
Control parameters & $a=10/t_f^2$ and $b=-2/3 t_f$
\end{tabular}
\end{ruledtabular}
\end{table}

\begin{table}
\caption{Typical protocols for state-to-state transformations in atomic units. All procedures were performed employing a thermal bath with temperature $T=5$\label{table:protocols}. $\bar{\Omega}_i$ and $\bar{\Omega}_f$ are the initial and final generalized Rabi frequencies of the TLS. The initial and final Gibbs state temperatures are $T_i$ and $T_f$.  }
\begin{ruledtabular}
\begin{tabular}{cccccc}
 Label& $\bar{\Omega}_i$ & $\bar{\Omega}_f$& $T_i$& $T_f$ 
\\
\hline
PE (Expansion) & 5 & 12 & 5 & 5 \\ PC (Compression) & 12 & 5 & 5 & 5\\ PE1 & 5 & 12 & 15 & 5\\
PE2  & 5 & 12 & 4 & 5\\ PEC  & 5 & 12 & 5 & 4\\
\end{tabular}
\end{ruledtabular}
\end{table}

%

\end{document}